\documentclass[final]{svjour2}
\usepackage{graphicx}
\usepackage{rotating}
\usepackage{amssymb}
\usepackage{mathptmx}
\DeclareMathAlphabet{\mathcal}{OMS}{cmsy}{m}{n}
\usepackage[numbers]{natbib}
\usepackage{amsmath}
\makeatletter
\journalname{Journal of Low Temperature Physics}

\bibpunct{}{}{,}{s}{}{,}

\begin{document}

\newcommand{\hdblarrow}{H\makebox[0.9ex][l]{$\downdownarrows$}-}
\title{Elementary excitations of antiferromagnetic spin-1 bosons in an optical lattice}

\author{M. Shinozaki$^1$ \and S. Tsuchiya$^2$ \and S. Abe$^2$ \and \\ T. Ozaki$^2$  \and T. Nikuni $^2$}

\institute{1:Department of Basic Science, The University of Tokyo, Komaba 3-8-1, Meguro-ku, Tokyo 153-8902, Japan\\
\email{shinozaki@vortex.c.u-tokyo.ac.jp}
\\2: Department of Physics, Tokyo University of Science, Kagurazaka 1-3, Shinjuku-ku, Tokyo 162-8601, Japan}

\date{\today}

\maketitle


\begin{abstract}

We study elementary excitations of spin-1 bosons with antiferromagnetic
 interaction in an optical lattice by applying the Gutzwiller
 approximation to the spin-1 Bose-Hubbard model. There appear various
 excitations associated with spin degrees of freedom in the
 Mott-insulator (MI) phase as well as in the superfluid (SF) phase. In
 this system, the ground state in the MI phase is known to exhibit a
 remarkable effect of even-odd parity of particle filling, in which even
 fillings stabilize the MI state due to formation of spin-singlet
 pairs. We find that excitation spectra in the MI phase exhibit
 characteristic features that reflect the even-odd parity effect of the
 ground state. We clarify evolution of elementary excitations across the
 quantum critical point of the SF-MI transition.

\end{abstract}

\section{Introduction}
The system of ultracold atomic gases in an optical lattice has attracted
attention since the observation of the quantum phase transition between
the superfluid (SF) and Mott-Insulator (MI) phases in
2002\cite{Greiner}. This system permits us to simulate various lattice
models including the Hubbard model in a perfect crystal with high degree of experimental
controllability and thus provides a new framework for the study of
quantum many-body systems\cite{Lewenstein}.
\par	 
Since the realization of Bose-Einstein condensation of spin-1 bosons\cite{Stenger}, there has been considerable interest in the system of bosonic atoms with spin degrees of freedom\cite{review}.
Although the system of spin-1 bosons trapped in an optical lattice has not been realized yet, previous theoretical studies predicted the existence of unique quantum phases and their remarkable features\cite{Lewenstein,Demler,Imambekov,Tsuchiya,Kimura}. 
In the case of spin-1 bosons with antiferromagnetic interaction, for example, the MI phase exhibits a remarkable even-odd parity effect in which the MI state with even fillings is stabilized because all particles form spin-singlet pairs, while in the MI phase with odd fillings there is an excess particle per site that cannot form singlet pairs\cite{Tsuchiya}. 
It has been also shown, using the Gutzwiller variational wave function,
that the phase transition between the SF and MI phases is of first
order around the tip of each Mott lobe, which is in sharp contrast with
the second order SF-MI transition for spinless bosons in an optical lattice\cite{Kimura}.
\par	
Previous studies of spin-1 bosons in an optical lattice mainly focused
on the ground state properties. In order to gain deeper understanding of
the system, it is useful to study elementary excitations. In the case of
spinless bosons, elementary excitations in both the SF and MI phases
have been studied theoretically\cite{Konabe, Ohashi, Altman, Huber} and
the evolution of elementary excitations across the quantum critical
point (QCP) of the SF-MI transition has been observed experimentally\cite{Ernst,Bissbort,Endres}.
\par
 In the present paper, we calculate the elementary excitations of antiferromagnetic spin-1 bosons in an optical lattice and analyze the excitation spectra around SF-MI phase transition point.
 In particular, we focus on the effect of the even-odd parity of the MI phase on the properties of elementary excitations.

\section{Formalism}
We consider the system of spin-1 bosons trapped in a two-dimensional
square optical lattice. In the case of a sufficiently deep optical
lattice, in which the tight-binding approximation is valid, the system
is described by the spin-1 Bose-Hubbard Hamiltonian\cite{Imambekov,Tsuchiya, Kimura}:
\begin{equation}
H=-J\sum_{\langle i,j\rangle,\alpha}(\hat{b}_{i,\alpha}^\dagger \hat{b}_{j,\alpha} + {\rm h.c.})
- \mu\sum_i\hat{n}_i
+ \frac{U_0}{2}\sum_i \hat{n}_i(\hat{n}_i -1)
+ \frac{U_2}{2}\sum_i (\hat{{\bf S}}_i^2 - 2\hat{n}_i),
\label{eq:Hamiltonian}
\end{equation}
where $\hat{b}_{i,\alpha}$ is the annihilation operator for an atom in
hyperfine state $|S=1,m=\alpha \rangle$ ($\alpha=1,0,-1$) on site
$i$. $\langle i,j\rangle$ denotes a summation over nearest-neighbor
sites. $J$ is the hopping matrix element, $\mu$ is the chemical
potential, and  $U_0$ and $U_2$ are  on-site spin-independent  and
spin-dependent interaction, respectively.
In this paper, we assume the antiferromagnetic interaction, i.e., $U_2>0$.
The operators $\hat{n}_i=\sum_\alpha \hat{n}_{i,\alpha}=\sum_\alpha \hat
b_{i,\alpha}^\dagger \hat b_{i,\alpha}$, and $\hat{{\bf S}}_i = \sum_{\alpha,\beta}\hat{b}_{i,\alpha}^\dagger ({\bf F})_{\alpha,\beta} \hat{b}_{i,\beta}$ represent the number of particles (particle filling) and the spin at site $i$, respectively.
Here, ${\bf F}$ denotes the spin-1 matrices.
\par	
For later use, we denote the local eigenstate of the spin and the particle number as $|S; N \rangle_i$ which satisfies 
$\hat{\bf S}_i^2 |S; N \rangle_i = S(S+1) |S; N \rangle_i$ and $\hat{n_i} |S; N \rangle_i = N |S; N \rangle_i$.
The SF phase is characterized by the order parameter $\Phi_\alpha = \langle \hat{b}_{\alpha} \rangle$.
\par
We assume the Gutzwiller-type factorized trial wave-function of the form
\begin{equation}
|\Psi(t) \rangle = \prod_i
 \sum_{n_1}\sum_{n_0}\sum_{n_{-1}}f_{n_1,n_0,n_{-1}} ^{( i)}(t)
 |n_1,n_0,n_{-1}\rangle_i\ .
\label{eq:GW}
\end{equation}
Here, $|n_1,n_0,n_{-1}\rangle_i$ is the local Fock state, $n_\alpha$ being the number of particle at site $i$ with magnetic quantum number $\alpha (=1,0,-1)$.
The time-dependent factor $f_{n_1,n_0,n_{-1}} ^{(i)}(t) \equiv f_{\bf n}^{(i)}(t)$ (hereafter we use the notation ${\bf n} \equiv (n_1,n_0,n_{-1})$) must satisfy the normalization condition $\sum_{\bf n} |f_{\bf n} ^{(i)} |^2=1$.
\par
The Gutzwiller approximation (\ref{eq:GW}), which is equivalent to the
perturbative mean-field approximation\cite{Tsuchiya,Yamamoto}, should do
well in the deep SF and MI regimes where mean field description is valid
due to small quantum fluctuation. In the vicinity of the QCP of the
SF-MI transition, although the Gutzwiller approximation is not
quantitatively correct, excitations in this regime can be qualitatively
captured by the Gutzwiller approximation as confirmed by the detection
of the amplitude mode near the QCP\cite{Endres}. This allows us
to deduce that our calculations are still useful for understanding the
characters of the excitations even in the transition region.
\par
In terms of the Fourier component $f_{\bf n}^{(i)} = \frac{1}{\sqrt{M}}\sum_{\bf k}f_{\bf n,k}e^{i{\bf k\cdot r}_i}$, where $M$ is the total number of the lattice sites, the effective action can be written as 
\begin{align}
	{\cal S}=\int dt \left[ i\hbar\sum_{\bf n} \sum_{\bf k} f_{\bf n,k}^* \frac{d}{dt} f_{\bf n,k} -E[\{f_{\bf n,k} \}] \right].
\end{align}
where $E[\{f_{\bf n,k}\}] = \langle \Psi|H|\Psi \rangle$ is the
variational energy function given in terms of the coefficients $\{
f_{\bf n,k} \}$
\begin{eqnarray}
E[\{f_{\bf n,k}\}]
&=& -Jz\sum_\alpha|\Phi_\alpha|^2+\sum_{\bf n}
\left\{	(U_0-U_2)n_1n_{-1}+(U_0+U_2)(n_1n_0+n_0n_{-1})-\mu N
\right. \nonumber\\
&&\left.+\frac{U_0}{2}n_0(n_0-1)+\frac{U_0+U_2}{2} [n_1(n_1-1) + n_{-1}(n_{-1}-1)] 
\right\}|f_{\bf n,k}|^2\nonumber\\
+&U_2&\sum_{\bf n}
	\left(
		\sqrt{n_0(n_0-1)}\sqrt{n_1+1}\sqrt{n_{-1}+1} f_{\bf
		n,k}^*f_{n_1+1,n_0-2,n_{-1}+1,{\bf k}}+{\rm c.c.}
	\right).
\end{eqnarray}
The variation of the action is given by 
\begin{equation}
\delta {\cal S}=
  \int dt \sum_{{\bf n},{\bf k}}
\left[ i\hbar \left(\delta f_{\bf n,k}^*\frac{d}{dt}f_{\bf n,k} - \delta f_{\bf n,k}\frac{d}{dt} f_{\bf n,k}^*\right)
  -\frac{\partial E}{\partial f_{\bf n,k}}\delta f_{\bf n,k} -
  \frac{\partial E}{\partial f_{\bf n,k}^*}\delta f_{\bf n,k}^*
\right]\ .
\end{equation}
Imposing the condition that $\delta {\cal S} = 0$ for an arbitrary $\delta f_{\bf n,k}$ leads to the equation of motion
\begin{equation}
i\hbar \frac{d}{dt}f_{\bf n,k} = \frac{\partial E}{\partial
 f_{\bf n,k}^*}\ .
\label{eq:motion}
\end{equation}
For the ground state, we assume the spatially uniform stationary solution $f_{\bf n,k} = \delta _{\bf k,0} \tilde f_{\bf n,k} e^{- i \tilde\omega t} $ which satisfies $\hbar \tilde \omega \tilde f_{\bf n,0} = \partial E / \partial \tilde f_{\bf n,0}^*$.
\par
In order to consider excitations, we assume small amplitude oscillations of the form:
\begin{equation}
f_{\bf n,k}(t) = [\delta_{\bf k,0} \tilde f_{\bf n,k} + \delta f_{\bf
 n,k}(t)] e^{-i\tilde\omega t }\ .
\label{eq:oscillation}
\end{equation}
Inserting Eq.~(\ref{eq:oscillation}) into Eq.~(\ref{eq:motion}) and linearizing in $\delta f$, we find 
\begin{equation}
 i\hbar \frac{d}{dt}  \delta f_{\bf n,k} =  \sum_{\bf n'}[X_{\bf
  n,n'}({\bf k}) \delta f_{\bf n',k} + Y_{\bf n,n'}({\bf k})
  \delta f_{\bf n',k}^*]\ ,
 \label{eq:linear}
\end{equation}
where the coefficient matrices $X$ and $Y$ are given by
\begin{equation}
	X_{\bf n,n'} ({\bf k}) = - \hbar\tilde\omega\delta_{\bf n,n'} +  \frac{\partial^2 E}{\partial f_{\bf n,k}^*\partial f_{\bf n',k}}, \ \ 
	Y_{\bf n,n'} ({\bf k}) =  \frac{\partial^2 E}{\partial f_{\bf
	n,k}^*\partial f_{\bf n',k}^*}\ .
	\label{eq:XY}
\end{equation}
It should be understood that the second derivative of the energy function in Eq.~({\ref{eq:XY}}) are evaluated with the stationary solution. In order to solve the above linearized equation, we assume the fluctuation in the form $\delta f_{\bf n,k}(t)=u_{\bf n,k}e^{-i\omega t}+v_{\bf n,k}^*e^{i\omega t}$. Inserting this into Eq.~(\ref{eq:linear}), we obtain a coupled linear algebraic equations
\begin{align}
	&\hbar\omega u_{\bf n,k} = \sum_{\bf n'}	[X_{\bf n,n'}({\bf k} )u_{\bf n',k} +Y_{\bf n,n'}({\bf k}) v_{\bf n',k}],
	\label{eq:Bogo1}
	\\
	&\hbar\omega v_{\bf n,k} = - \sum_{\bf n'} [X_{\bf n,n'}^*({\bf k}) v_{\bf n',k} +Y_{\bf n,n'}^*({\bf k}) u_{\bf n',k}].
	\label{eq:Bogo2}
\end{align}
We calculate the excitation spectrum $\hbar\omega$ by diagonalizing Eqs.~(\ref{eq:Bogo1}) and (\ref{eq:Bogo2}).
To study the characteristics of elementary excitations, we examine eigenvectors\\ $(u_{\bf n,k}, v_{\bf n,k})$.
\section{Results}
We discuss the characteristic behavior of the elementary excitations in the MI and SF phases and their evolution across the QCP of the SF-MI transition. 
Our main focus is on the distinct nature of elementary excitations in
the MI phases with even and odd fillings and their evolution across the
QCP that reflects the even-odd parity effect in the ground state.
\begin{figure}[htb]
\begin{center}
\includegraphics[width=1\linewidth,keepaspectratio]{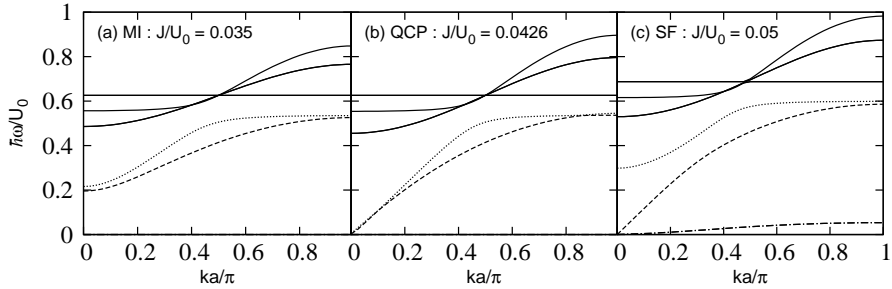}
\end{center}
\caption{Excitation spectrum for (a) $J/U_0=0.035$ (MI phase), (b) $J/U_0=0.0426$ (QCP), and (c) $J/U_0=0.05$ (SF phase). 
We set $U_2/U_0 = 0.04$ and tune the chemical potential so that the particle filling is set to unity $(N=1)$.
The lowest branch (dashed line) and second lowest branch (dotted line)
 in (a) and (b) are hole and particle type excitation in the MI phase,
 respectively. They become the gapless NG mode and the gapful Higgs mode in the SF phase.
The lowest branch in the SF phase in (c) (dash-dot line) is the doubly degenerate spin wave mode.	
}
\label{n=1}
\end{figure}
\par
Figure~\ref{n=1} shows the evolution of elementary excitations across the QCP between the MI and SF phases for $N=1$. 
We set $U_2/U_0=0.04$ assuming $^{23}{\rm Na}$\cite{Tsuchiya}. 
Analogous to the case of spinless bosons\cite{Konabe,Ohashi}, the typical elementary excitations in the MI phase of spin-1 bosons have particle or hole character.
The lowest branch of the excitations in the MI phase in Fig.~\ref{n=1} (a) is a hole type excitation that involves the transition of the local Fock state $|S=1;N=1\rangle_i\to |S=0;N=0\rangle_i$, while the second lowest branch is a particle type excitation that involves the transition $|S=1;N=1\rangle_i\to |S=0;N=2\rangle_i$. 
Both the particle and hole type excitations carry no spin. 
A particle excitation involves formation of a local singlet pair on each lattice site due to the antiferromagnetic interaction.
Other upper branches are particle or hole excitations involving higher excited states of the local Fock states, e.g., the three upper branches in Fig.~\ref{n=1} (a) correspond to the translation $|S=1;N=1\rangle_i\to |S=2;N=2\rangle_i$.
\par
The energy of the lowest two branches decreases progressively as we approach the critical point from the MI side. Precisely at the critical point, both the particle and hole excitations soften at $({\bf k},\omega)=(0,0) $ and undergo Bose condensation. 
This leads to the transition to the SF phase.
The order parameter of the SF phase determined to minimize the
mean-field energy takes the form $(\Phi_1,\Phi_0,\Phi_{-1})\propto
(0,1,0)$, which shows that the ground state in the SF phase is a polar state\cite{Tsuchiya}.
\par
In Fig.~\ref{n=1}, the particle and hole excitations in MI state continuously evolve across the critical point and after the condensation they become the gapless Nambu-Goldstone (NG) mode and gapful Higgs mode\cite{Huber,Endres} that arise from phase and amplitude fluctuation of superfluid order parameter, respectively. 
In addition, the doubly degenerate gapless spin wave modes arise in the SF phase. 
They are degenerate with the ground state at the critical point.
We note that this behavior is an artifact of the Gutzwiller-type wave function where the MI state is described by a local Fock state and thus the spin correlation between different lattice sites is neglected.
If we take into account spin correlation in the MI phase, the spin wave modes would have dispersion in the MI phase.
\par
In the deep SF regime ($J/U_0\gg 1$), the NG mode and the spin wave mode
remain as low lying excitations, while the excitation energy of the
Higgs amplitude mode and other higher modes linearly increases as the
hopping amplitude $J$ increases. It becomes infinite in the SF limit
when $J/U_0$ goes infinite. 
In this regime, excitations with small wave vector $(k \ll \pi/a)$ are expected to be well described by the Gross-Pitaevskii (GP) equation for a spin-1 Bose condensate.
In fact, the GP analysis for the polar phase predicts two gapless modes: density mode and spin wave mode\cite{Ho}.
This is thus consistent with the present results.
\begin{figure}[htb]
\begin{center}
\includegraphics[width=1\linewidth,keepaspectratio]{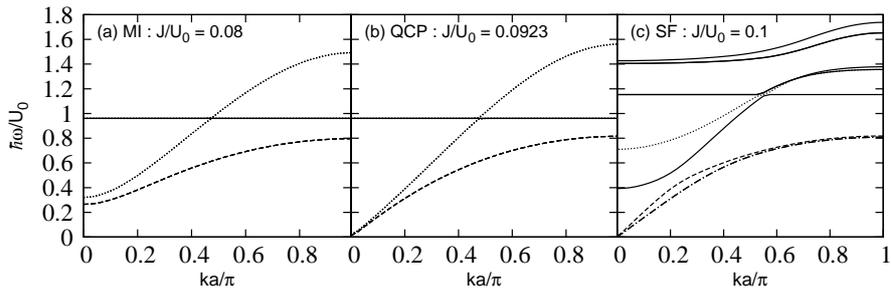}
\end{center}
\caption{Excitation spectrum for (a) $J/U_0=0.08$ (MI phase), (b) $J/U_0=0.0923$ (QCP), and (c) $J/U_0=0.1$ (SF phase).  
We set $U_2/U_0=0.32$ and tune the chemical potential so that the particle filling is set to $N=2$.
The lowest branch (dashed line) and second lowest branch (dotted line) in (a) and (b) are hole and particle type excitation in the MI phase, respectively.
They are both triply degenerate. They evolve into the gapless NG modes (dashed and dash-dot lines) and the gapful Higgs mode (solid and dotted lines) in the SF phase in (c).
}
\label{n=2}
\end{figure}
Figure~\ref {n=2} shows the evolution of elementary excitations across the QCP for $N=2$.
To compare with the results for $N=1$, we set $U_2/U_0 = 0.32$ for which the transition from the MI phase to the SF phase is of second order\cite{Kimura}.
The lowest  branch in Fig.~\ref{n=2} (a) is a hole type excitation that involves the transition of the local Fock state $|S=0;N=2\rangle_i\to |S=1;N=1\rangle_i$, while the second lowest branch is a particle type excitation involving the transition $|S=0;N=2 \rangle_i\to |S=1;N=3\rangle_i$.
They both carry spin 1 and therefore triply degenerate.
We note that the dispersionless branch is also an artifact of the approximation.
The spectrum becomes dispersionless whenever the change of the local wave function of the Gutzwiller-type wave function does not involve the hopping term.
\par
The lowest two excitations go down as approaching the critical point from the MI side. 
At the critical point they undergo condensation in Fig.~\ref{n=2} (b) and the system enters the SF phase.
\par
The triply degenerate  particle and hole excitations in the MI phase become three gapless modes and three gapful modes in the SF phase. 
One of the gapless modes has  large density component, while the other two modes have large spin component and degenerate.
They evolve in the deep SF regime into the pure density and spin wave modes of the GP equation\cite{Ho}.
\par
In Fig.~\ref{n=2} (c), the lowest gapful branch is a single mode and the second lowest gapful branch is doubly degenerate. 
They have complex character involving mixed fluctuation of order parameter amplitude and longitudinal spin fluctuation\cite{Shinozaki}.
\par
Remarkably, the elementary excitations in the MI phase with $N=1$ and $N=2$ exhibit distinct characteristic features. 
For example, a particle excitation for $N=1$ carries a local spin-singlet pair, while that for $N=2$ carries spin 1.
This clearly reflects the parity effect of the ground state in the MI phase. 
The phase transition to the SF phase can be associated with condensation of these excitations with different spin character. 
This strongly suggests that the SF phase in the vicinity of the critical point has different magnetic property beyond the mean-field analysis\cite{Tsuchiya, Kimura}.
Although the gapless NG modes in the SF phase has qualitatively the same character for $N=1,2$ that is consistent with the GP analysis in the deep SF regime, the gapful Higgs modes seem to have distinct features reflecting the different magnetic property of the SF ground state\cite{Shinozaki}.
\par
In the case of spinless bosons in an optical lattice, elementary
excitations have been probed in both the SF and MI phases
experimentally. In the MI phase, particle and hole type excitations have
been observed by tilting the lattice potential\cite{Greiner}. In the SF
phase,  gapless phonon mode has been observed by Bragg
spectroscopy\cite{Ernst}, and the amplitude mode in the SF phase has been
observed by shaking lattice potential\cite{Endres}. It should be possible to
apply the same kind of experimental techniques to spin-1 bosons to probe
excitations. However, the above experimental techniques cannot
distinguish excitations with different magnetic characters. To identify
the different magnetic characters of elementary excitations,
spin-selective optical lattice\cite{McKay} may be useful. For example, it may
be possible to probe particle and hole excitations with spin by shaking
a lattice potential that is coupled to a single spin component.

\section{Conclusions}
We have investigated elementary excitations of antiferromagnetically interacting spin-1 bosons in an optical lattice within the Gutzwiller approximation.
We calculated excitation spectra in the MI and SF phases and clarified the evolution of excitations across the QCP of the SF-MI transition. 
We found that particle and hole type excitations in the MI phase have distinct spin characters that reflect the even-odd parity effect in the ground state. 
These excitations undergo condensation at the critical point and become the collective modes involving fluctuation of order parameter and spin in the SF phase.
\begin{acknowledgements}
We thank D. Yamamoto, I. Danshita, M. Kunimi, Y. Kato, and T. Kimura for helpful discussions. S.T. was supported by Grant-in-Aid for Scientific Research, No. 24740276.
\end{acknowledgements}


\end{document}